\documentclass[aps,prd,preprint,groupedaddress,showpacs,amsmath,amssymb]{revtex4}
\usepackage{graphicx}
\usepackage{times,amsmath,amsfonts,amsthm}
\usepackage{mathrsfs}
\newtheorem{theorem}{Theorem}

\newcommand{\prop}[1]{\textit{\textbf {#1}}$'$.}
\begin{document}
\title{Topology and Closed Timelike Curves II: Causal structure}
\author{Hunter Monroe}
\affiliation{International Monetary Fund, Washington, DC 20431}
\begin{abstract}
Because no closed timelike curve (CTC) on a Lorentzian manifold can be deformed to a point, any such manifold containing a CTC must have a topological feature, to be called a timelike wormhole, that prevents the CTC from being deformed to a point. If all wormholes have horizons, which typically seems to be the case in space-times without exotic matter, then each CTC must transit some timelike wormhole's horizon. Therefore, a Lorentzian manifold containing a CTC may nevertheless be causally well behaving once its horizon's are deleted. For instance, there may be a Cauchy-like surface through which every timelike curve passes one and only once before crossing a horizon.
\end{abstract}
\date{\today}
\thanks{
In honor of the retirement from Davidson College of Dr. L. Richardson King, an extraordinary teacher and mathematician. Email: relpap at huntermonroe.com.  The views expressed in this paper are those of the author and do not necessarily represent those of the International Monetary Fund or IMF policy.}
\hyphenpenalty=3000
\maketitle
\section{Censorship of chronological violations}\label{Censorshipofchronologicalviolations}
An argument that CTCs should not be considered pathological is that all CTCs typically transit a timelike wormhole which prevent the CTC from being deformed to a point (see Monroe \cite{Monroe:2006b}), and all wormholes in a vacuum space-time typically have an event horizon. Such a CTC is then ``censored" by passing through the wormhole's event horizon and will be called a \textit{c-CTC}; other CTCs are \textit{uncensored-} or \textit{u-CTCs}.

The causal structure of a space-time ignoring c-CTCs can be examined by analyzing a space-time with points on event horizons excised. Modifying the definitions to this setting, most results Hawking and Ellis \cite{Hawking-Ellis} Chapter 6 carry over in some form, and the excised space-time may be well-behaved by modified criteria. For instance, a Cauchy-like surface may exist through which all timelike curves pass one and only once before crossing an event horizon. A few results do not carry over: the proof of Proposition 6.4.2, stating that compact space-times contain CTCs, cannot be modified to show these are u-CTCs. The Appendix presents restatements of Chapter 6's definitions and results that are not needed for this section.

Consider a space-time $(\mathscr{M},\textbf{g})$. Define $\mathscr{M}_c=\mathscr{M}-\dot{J}^-(\mathscr{I}^+,\overline{\mathscr{M}})-\dot{J}^+(\mathscr{I}^-,\overline{\mathscr{M}})$, that is, excise all points on event horizons. Below, the causal structure of a space-time $\mathscr{M}$ containing c-CTCs is considered by analyzing $\mathscr{M}_c$.

Say a time-orientable space-time $\mathscr{M}$ satisfying the Einstein vacuum equation is \textit{safe} if all wormholes have event horizons; this is a weak assumption in a vacuum assumption in the absence of exotic matter (Morris and Thorne \cite{Morris:1988cz}). This section and the next will establish that safe space-times may contain c-CTCs cannot contain no u-CTCs and are therefore well behaved by standard criteria with appropriate modifications.

\begin{theorem}\label{trivialtopology}
Suppose space-time $\mathscr{M}$ is safe. Then, $\mathscr{M}_c$ is simply connected.
\end{theorem}
\begin{proof}
In the absence of topological defects, every CTC passes through a wormhole (see Monroe \cite{Monroe:2006b}). Every wormhole of $\mathscr{M}$ has an event horizon in a safe space-time, so the space-time $\mathscr{M}$ becomes simply connected when its event horizons are excised to create $\mathscr{M}_c$.
\end{proof}

Suppose $\mathscr{M}$ is time orientable. A \textit{u-curve} is any curve of non-zero extent on $\mathscr{M}_c$. For sets $\mathscr{S}$ and $\mathscr{U}$, the \textit{c-chronological future} $I^+_c(\mathscr{S},\mathscr{U})$ \textit{of} $\mathscr{S}$ \textit{relative to} $\mathscr{U}$ is the set of all points in $\mathscr{U}$ which can be reached from $\mathscr{S}$ by a future-directed timelike u-curve. $I^+_c(\mathscr{S},\mathscr{M}_c)$ will be denoted by $I^+_c(\mathscr{S})$. The \textit{c-causal future of} $\mathscr{S}$ \textit{relative to} $\mathscr{U}$, denoted by $J^+_c(\mathscr{S},\mathscr{U})$, is defined similarly.

A point $p$ is a \textit{c-future endpoint} of a future-directed non-spacelike u-curve $\gamma:F\rightarrow\mathscr{M}_c$ if for every neighborhood $\mathscr{V}$ of $p$ there is a $t\in{}F$ such that $\gamma(t_1)\in\mathscr{V}$ for every $t_1\in{}F$ with $t_1\geq{}t$. A non-spacelike u-curve is \textit{c-future-inextendible} (respectively \textit{c-future-inextendible in a set} $\mathscr{S}$) if it has no future endpoint in $\mathscr{M}_c$ (respectively \textit{c-future-inextendible in a set} $\mathscr{S}$).

A set $\mathscr{S}$ is \textit{c-achronal} if $I^+_c(\mathscr{S})\cap\mathscr{S}$ is empty, in other words, if there are no two points of $\mathscr{S}$ that lie on a timelike u-curve. $\mathscr{S}$ is a \textit{c-future set} if $\mathscr{S}\supset{}I^+_c(\mathscr{S})$. An example of a result from Hawking and Ellis that carries over with the modified causality definitions above is:

\prop{Proposition 6.3.1} If $\mathscr{S}$ is a c-future set then $\dot{\mathscr{S}}$, the boundary of $\mathscr{S}$ in $\mathscr{M}_c$ is a closed, imbedded, c-achronal three-dimensional $C^{1-}$ manifold.

A set with the properties of $\dot{\mathscr{S}}$ in Proposition 6.3.1$'$ a \textit{c-achronal boundary}. An open set $\mathscr{U}$ is \textit{c-causally simple} if for every compact set $\mathscr{K}\subset\mathscr{U}$, $\dot{J}^+_c(\mathscr{K})\cap\mathscr{U}=E^+_c(\mathscr{K})\cap\mathscr{U}$ and $\dot{J}^-_c(\mathscr{K})\cap\mathscr{U}=E^-_c(\mathscr{K})\cap\mathscr{U}$.

A space-time satisfies the \textit{c-chronology condition} if there are no u-CTCs. Trivially, the c-chronology condition is weaker than the chronology condition, because the latter rules out c-CTCs as well. The set of all points in $\mathscr{M}_c$ that lie on a u-CTC is called the \textit{c-chronology violating} set of $\mathscr{M}_c$. For a safe space-time $\mathscr{M}$, the c-chronology violating set of $\mathscr{M}_c$ is empty, by definition.

Proposition 6.4.2 is stated in its original form, because it cannot be modified to apply to $\mathscr{M}_c$:

\textit{\textbf{Proposition 6.4.2}} If $\mathscr{M}$ is compact, the chronology violating set of $\mathscr{M}$ is non-empty.

Hawking and Ellis point out that any compact, four-dimensional manifold on which there is a Lorentzian metric cannot be simply connected. They conclude from this property and Proposition 6.4.2 that it is reasonable to assume space-time is compact. By contrast, if c-CTCs are considered reasonable, it will be argued below that compactness is a desirable feature. If $\mathscr{M}$ is compact and therefore contains CTCs, the reasoning in Hawking and Ellis' proof of Proposition 6.4.2 does not necessarily imply that $\mathscr{M}_c$ contains CTCs if $\mathscr{M}$ contains a wormhole with event horizon---in that case $\mathscr{M}_c$ is not compact at the excised points on event horizons. Let $\mathscr{M}_c'$ be the compactification of $\mathscr{M}_c$; in other words, event horizons of $\mathscr{M}$ are unidentified, and $\mathscr{M}_c'$ remains simply connected (see Theorem \ref{trivialtopology}). Although $\mathscr{M}_c'$ is compact, the proof cannot assume $\mathscr{M}_c'$ can be covered by open sets of the form $I^+_c(q)$. For a given wormhole $w$, label the unidentified horizons $\mathscr{I}^-_w$ and $\mathscr{I}^+_w$ respectively; this notation will be justified by Theorem \ref{causalboundarytheorem}.

\begin{theorem}
$\mathscr{M}_c'$ cannot be covered by open sets of the form $I^+_c(q)$ if $\mathscr{M}$ contains a wormhole with event horizon. If $\mathscr{M}_c'$ is covered by open c-future sets, these have a finite subcover.
\end{theorem}
\begin{proof}
For any point $q\in\mathscr{I}^-_w$, $q\notin{}I^+_w(q)$. If $\mathscr{M}_c'$ is covered by open c-future sets, some sets must include a non-zero measure of $\mathscr{I}^-_w$, so there is a finite subcover.
\end{proof}
The \textit{strong c-causality condition} is said to hold at $p$ if every neighborhood of $p$ contains a neighborhood of $p$ which no non-spacelike u-curve intersects more than once.

The region $D^+_c(\mathscr{S})$ to the future of $\mathscr{S}$ is called the \textit{c-future Cauchy development} or \textit{c-domain of dependence} of $\mathscr{S}$, defined as the set of all points $p\in\mathscr{M}_c$ such that every c-past-inextendible non-spacelike u-curve through $p$ intersects $\mathscr{S}$. Appropriate data on a closed set $\mathscr{S}$ would determine events in not only in $\mathscr{M}_c$ but also in $\mathscr{M}$, and there is an the additional consistency condition in the latter case.

The future boundary of $D^+_c(\mathscr{S})$, that is $\overline{D^+_c(\mathscr{S})}-I^-(D^+_c\mathscr{S})$, marks the limit of the region that can be predicted from knowledge of data on $\mathscr{S}$. Call this closed c-achronal set the \textit{c-future Cauchy horizon} of $\mathscr{S}$ and denote it by $H^+_c(\mathscr{S})$. Define the c-edge$(\mathscr{S})$ for a c-achronal set $\mathscr{S}$ as the set of all points $q\in\overline{\mathscr{S}}$ such that in every neighborhood $\mathscr{U}$ of $q$ there are points $p\in{}I^-(q,\mathscr{U})$ and $r\in{}I^+(q,\mathscr{U})$ which can be joined by a timelike u-curve in $\mathscr{U}$ which does not intersect $\mathscr{S}$.

\prop{Corollary to Proposition 6.5.3} If c-edge$(\mathscr{S})$ vanishes, then $H^+_c(\mathscr{S})$, if non-empty, is a c-achronal three-dimensional imbedded $C^{1-}$ manifold which is generated by null geodesic segments which have no past endpoint in $\mathscr{M}_c$.

Call such a c-acausal set $\mathscr{S}$ with no c-edge a \textit{partial c-Cauchy surface}. That is, a partial c-Cauchy surface is a spacelike hypersurface which no non-spacelike u-curve intersects more than once. Define $D_c(\mathscr{S})=D^+_c(\mathscr{S})\cup{}D^-_c(\mathscr{S})$. A partial c-Cauchy surface $\mathscr{S}$ is a global c-Cauchy surface (or simply \textit{c-Cauchy surface}) if $D_c(\mathscr{S})$ equals $\mathscr{M}_c$. That is, a c-Cauchy surface is a spacelike hypersurface which every non-spacelike u-curve intersects once.

Friedman \cite{Friedman:2004jr} notes that the initial value problem is well-defined on a class of space-times broader than those which are globally hyperbolic. It is shown below that this larger class includes c-globally hyperbolic space-times.

Suppose one is given a three-dimensional manifold $\mathscr{S}$ with certain initial data $\omega$ on it. The \textit{c-Cauchy problem} requires one to find a four-dimensional manifold $\mathscr{M}$ (without subscript), an imbedding $\theta:\mathscr{S}\rightarrow\mathscr{M}$ and a metric \textbf{g} on $\mathscr{M}$ which satisfies the Einstein equations, agrees with the initial values on $\theta(\mathscr{S})$ and is such that $\theta(\mathscr{S})$ is an c-Cauchy surface for $\mathscr{M}$. Note that the c-Cauchy problems for $\mathscr{M}$ and $\mathscr{M}_c$ are distinct; in the former case, there are additional consistency constraints between $\mathscr{I}^+_w$ and $\mathscr{I}^-_w$.

\begin{theorem}
For a safe space-time $\mathscr{M}$, initial data on a c-Cauchy surface $\mathscr{S}_c$ is sufficient to predict not only $\mathscr{M}_c$ but also $\mathscr{M}_c'$.
\end{theorem}
\begin{proof}
By the definition of c-Cauchy surface, $D_c(\mathscr{S}_c)$ equals $\mathscr{M}_c$. The event horizons of $\mathscr{M}_c$ are limit points of $D_c(\mathscr{S}_c)$. Therefore, the Cauchy development includes the event horizons.
\end{proof}

Say a safe space-time $\mathscr{M}$ is \textit{Novikov consistent} with respect to c-Cauchy surface $\mathscr{S}$ if the c-Cauchy development of $\mathscr{S}$ in $\mathscr{M}_c$ coincides with the Cauchy development of $\mathscr{S}$ in $\mathscr{M}$. It is not obvious that Novikov consistency holds between $\mathscr{I}^-_w$ and $\mathscr{I}^+_w$.


A set $\mathscr{N}$ is \textit{c-globally hyperbolic} if the strong c-causality assumption holds on $\mathscr{N}$ and if for any two points $p,q\in\mathscr{N}$, $J_c^+(p)\cap{}J_c^-(q)$ is compact and contained in $\mathscr{N}$.

\prop{Proposition 6.6.8} If an open set $\mathscr{N}$ is c-globally hyperbolic, then $\mathscr{N}$, regarded as a manifold, is homeomorphic to $R^1\times\mathscr{S}$ where $\mathscr{S}$ is a three-dimensional manifold, and for each $a\in{}R^1$, $a\times\mathscr{S}$ is an c-Cauchy surface for $\mathscr{N}$.

Theorem \ref{trivialtopology} shows that $\mathscr{M}_c$ is simply connected, so there is no topological obstacle to $\mathscr{M}_c$ being c-globally hyperbolic.
%
\section{Causal completeness}\label{Causalcompleteness}
It is shown below that $\mathscr{I}^-_w$ and $\mathscr{I}^+_w$ are causal boundaries of $\mathscr{M}_c$. Assume $(\mathscr{M}_c,\textbf{g})$ satisfies the strong c-causality condition. Define an \textit{c-indecomposable past set}, abbreviated as c-IP, as a set that is with the properties (1) open; (2) a c-past set; and (3) cannot be expressed as the union of two proper subsets which have the properties (1) and (2). Divide c-IPs into two classes: \textit{c-proper IPs (c-PIPs)} which are the pasts of points in $\mathscr{M}_c$, and \textit{c-terminal IPs (c-TIPs)} which are not the past of any point in $\mathscr{M}_c$. Then c-TIPs and c-TIFs represent the \textit{causal boundary} of $(\mathscr{M}_c,\textbf{g})$.
\begin{theorem}\label{causalboundarytheorem}
The causal boundary of $\mathscr{M}_c$ includes $\mathscr{I}^-_w$ and $\mathscr{I}^+_w$ for each wormhole of $\mathscr{M}_c$.
\end{theorem}
\begin{proof}
Let $p\in\mathscr{I}^+_w\subset\mathscr{M}$. The set $I^-(p)\cap\mathscr{M}_c$ is a c-TIP.
\end{proof}

A drawback of assessing the completeness of space-time using geodesic completeness is that this is not a conformally invariant property: Whether a curve is a geodesic and whether its parameter is affine are not conformally invariant properties. A conformally invariant method of characterizing whether space-time is complete is to say that a space-time $(\mathscr{M},\textbf{g})$ is \textit{causally complete} if it has no causal boundary (causal boundaries are conformally invariant). A causally complete space-time is trivially geodesically and b-complete.

No spherical space-time can be causally complete; see the Corollary to Proposition 6.4.2 above. Furthermore, no asymptotically flat space-time is causally complete, unless $\mathscr{I}^+$ and $\mathscr{I}^-$ are identified so these are no longer causal boundaries. \footnote{Take the plane of simultaneity of an inertial observer, and consider how it evolves over time. In a space-time in which the vacuum Einstein equations hold universally, the stress energy tensor is identically zero, the trace of the Riemann tensor is zero, and the initial effects of tidal curvature to the first order are volume preserving. Map time dilation $\frac{1}{\sqrt{1-v^2}}$ into a relative velocity $v$ (see Monroe \cite{Monroe:2005gq}). The distortion of the observer's plane of simultaneity over time can be seen as a volume-preserving flow; causal completeness rules out the flow of volume to or from asymptotic infinity.}

A causally complete asymptotically flat space-time is conformally equivalent to a compact space-time. Therefore, among the set of causally complete space-times, there is no advantage of asymptotically flat over compact space-times. Thus, the two arguments against compact space-times offered by Hawking and Ellis---the existence of chronology violations (Proposition 6.4.2) and any four dimensional manifold with a Lorentzian metric cannot be simply connected---are inherent features of causally complete space-times. They point out that a compact space-time is really a non-compact manifold in which points have been identified, and suggest it would seem physically reasonable not to identify points but to regard the covering manifold as representing space-time. Not identifying $\mathscr{I}^+$ and $\mathscr{I}^-$ destroys causal completeness.
\begin{theorem}
A space-time $(\mathscr{M},\textbf{g})$ is causally complete if and only if there is a CTC through every point.
\end{theorem}
\begin{proof}
$\rightarrow$: A causally complete space-time $\mathscr{M}$ is conformally equivalent to a compact space-time. By Proposition 6.4.2, $\mathscr{M}$ contains CTCs.

$\leftarrow$: Suppose $\mathscr{M}$ is not causally complete, and let $p$ be a point a causal boundary. There is no CTC through $p$, which is a contradiction.
\end{proof}
Compact space-times for which there is a Lorentz metric are causally complete.
\section{Restatements of other causality results}
This appendix states modified versions of other results in Hawking and Ellis \cite{Hawking-Ellis} not needed for the line of argument above.

$I^+_c(\mathscr{S})$ is an open set, since if $p\in\mathscr{M}_c$ can be reached by a future-directed timelike u-curve from $\mathscr{S}$ then there is a small neighborhood of $p$ which can be so reached.

Then $\overline{I^+_c}(p,\mathscr{U})=\overline{J^+_c}(p,\mathscr{U})$ and $\dot{I}^+_c(p,\mathscr{U})=\dot{J}^+_c(p,\mathscr{U})$ where for any set $\mathscr{K}$, $\overline{\mathscr{K}}$ denotes the closure of $\mathscr{K}$ in $\mathscr{M}_c$ and $\dot{\mathscr{K}}\equiv\overline{\mathscr{K}}\cap(\overline{\mathscr{M}_c-\mathscr{K}})$ denotes the boundary of $\mathscr{K}$ in $\mathscr{M}_c$.

The \textit{c-future horismos of $\mathscr{S}$ relative to $\mathscr{U}$}, denoted by $E^+_c(\mathscr{S},\mathscr{U})$, is defined as $$J^+_c(\mathscr{S},\mathscr{U})-I^+_c(\mathscr{S},\mathscr{U});$$ write $E^+_c(\mathscr{S})$ for $E^+_c(\mathscr{S},\mathscr{M}_c)$.

\prop{Lemma 6.2.1} Let $\mathscr{S}$ be an open set in $\mathscr{M}_c$ and let $\lambda_n$ be an infinite sequence of non-spacelike curves in $\mathscr{S}$ (which are u-curves by definition) which are future-inextendible in $\mathscr{S}$. If $p\in\mathscr{S}$ is a limit point of $\lambda_n$, then through $p$ there is a non-spacelike curve $\lambda$ which is future-inextendible in $\mathscr{S}$ and which is a limit curve of $\lambda_n$.

\prop{Proposition 6.4.1} The c-chronology violating set of $\mathscr{M}_c$ is the disjoint union of sets of the form $I^+_c(q)\cap{}I^-_c(q)$, $q\in\mathscr{M}_c$.

Say that the \textit{c-causality condition} holds if there are no closed non-spacelike u-curves.

\prop{Proposition 6.4.3} The set of points at which the c-causality condition does not hold is the disjoint union of sets of the form $J^+_c(q)\cap{}J^-_c(q)$, $q\in\mathscr{M}_c$.

\prop{Proposition 6.4.4} If $\gamma$ is a closed null geodesic u-curve which is incomplete in the future direction then there is a variation of $\gamma$ which moves each point of $\gamma$ towards the future and which yields a closed timelike u-curve.

\prop{Proposition 6.4.5} If

(\textit{a}) $R_{ab}K^a K^b\geq0$ for every null vector $\textbf{K}$;

(\textit{b}) the generic condition holds, i.e. every null geodesic contains a point at which $K_{[a}R_{b]cd[e}K_{f]}K^c K^d$ is non-zero, where $\textbf{K}$ is the tangent vector;

(\textit{c}) the c-chronology condition holds on $\mathscr{M}_c$,

then the c-causality condition holds on $\mathscr{M}_c$.

The \textit{c-future} \textit{distinguishing condition} is said to hold at $p\in\mathscr{M}_c$ if every neighborhood of $p$ contains a neighborhood of $p$ which no future directed non-spacelike u-curve from $p$ intersects more than once.

\prop{Proposition 6.4.6} If conditions (\textit{a}) to (\textit{c}) of Proposition 6.4.5 hold and if in addition (\textit{d}) $\mathscr{M}_c$ is null geodesically complete except at event horizons, then the strong c-causality condition holds on $\mathscr{M}_c$.

A non-spacelike u-curve $\gamma$ that is c-future-inextendible can do one of three things as one follows it to the future: it can:

(i) enter and remain within a compact set $\mathscr{S}$, in which case it is \textit{totally c-future imprisoned},

(ii) not remain within any compact set but continually re-enter a compact set $\mathscr{S}$, in which case it is \textit{partially c-future imprisoned},

(iii) not remain within any compact set $\mathscr{S}$ and not re-enter any such set more than a finite number of times.

\prop{Proposition 6.4.7} If the strong c-causality condition holds on a compact set $\mathscr{S}$, there can be no c-future-inextendible non-spacelike u-curve totally or partially c-future imprisoned in $\mathscr{S}$.

\prop{Proposition 6.4.8} If the future or past distinguishing condition holds on a compact set $\mathscr{S}$, there can be no c-future-inextendible non-spacelike u-curve totally c-future imprisoned in $\mathscr{S}$.

Say that the \textit{stable c-causality condition} holds on $\mathscr{M}_c$ if the space-time metric \textbf{g} has an open neighborhood in the $C^0$ open topology such that there are no u-CTCs in any metric belonging to the neighborhood.

\prop{Proposition 6.4.9} The stable c-causality condition holds everywhere on $\mathscr{M}_c$ if and only if there is a function $f$ on $\mathscr{M}_c$ whose gradient is everywhere timelike.

Let $\tilde{D}_c(\mathscr{S})$ be the set of all points $p\in\mathscr{M}_c$ such that every c-past-inextendible timelike u-curve through $p$ intersects $\mathscr{S}$. Then:

\prop{Proposition 6.5.1} $\tilde{D}_c(\mathscr{S})=\overline{D^+_c}(\mathscr{S})$.

By an argument similar to that in Proposition 6.3.1 it follows that if c-edge($\mathscr{S}$) is empty for a non-empty c-achronal set $\mathscr{S}$, then $\mathscr{S}$ is a three-dimensional imbedded $C^{1-}$ submanifold.

\prop{Proposition 6.5.2} For a closed c-achronal set $\mathscr{S}$, c-edge$(H^+_c(\mathscr{S}))=$c-edge$(\mathscr{S})$.

\prop{Proposition 6.5.3} Let $\mathscr{S}$ be a closed c-achronal set. Then $H^+_c(\mathscr{S})$ is generated by null geodesic segments which either have no past endpoints in $\mathscr{M}_c$ or have past endpoints at c-edge$(\mathscr{S})$.

\prop{Proposition 6.6.1} An open c-globally hyperbolic set $\mathscr{N}$ is c-causally simple.

Following Leray, for points $p,q\in\mathscr{M}_c$ such that strong c-causality holds on $J_c^+(p)\cap{}J_c^-(q)$, define $C(p,q)$ to be the space of all (continuous) non-spacelike u-curves from $p$ to $q$, regarding two curves $\gamma(t)$ and $\lambda(u)$ as representing the same point of $C(p,q)$ if one is a reparameterization of the other.

\prop{Proposition 6.6.2} Let strong c-causality hold on an open set $\mathscr{N}$ such that $\mathscr{N}=J_c^-(\mathscr{N})\cap{}J_c^+(\mathscr{N})$. Then $\mathscr{N}$ is c-globally hyperbolic if and only if $C(p,q)$ is compact for all $p,q\in\mathscr{N}$.

\prop{Proposition 6.6.3} If $\mathscr{S}$ is a closed c-achronal set, then int$(D_c(\mathscr{S}))\equiv{}D_c(\mathscr{S})-\dot{D}_c(\mathscr{S})$, if non-empty, is c-globally hyperbolic.

\prop{Lemma 6.6.4} If $p\in{}D_c^+(\mathscr{S})-H_c^+(\mathscr{S})$, then every c-past-inextendible non-spacelike u-curve through $p$ intersects $I_c^-(\mathscr{S})$.

\prop{Lemma 6.6.5} The strong c-causality condition holds on int$(D_c(\mathscr{S}))$.

\prop{Lemma 6.6.6} If $q\in$int$(D_c(\mathscr{S}))$, then $J_c^+(\mathscr{S})\cap{}J_c^-(q)$ is compact or empty.

\prop{Proposition 6.6.7} If $\mathscr{S}$ is a closed c-achronal set such that $J^+_c(\mathscr{S})\cap{}J^-_c(\mathscr{S})$ is both strongly c-causal and either (1) c-acausal or (2) compact, then $D_c(\mathscr{S})$ is c-globally hyperbolic.

\prop{Proposition 6.7.1} Let $p$ and $q$ lie in an c-globally hyperbolic set $\mathscr{N}$ with $q\in{}J^+_c(p)$. Then there is a non-spacelike \textit{c-geodesic} (i.e. a geodesic which is an u-curve) from $p$ to $q$ whose length is greater than or equal to that of any other non-spacelike u-curve from $p$ to $q$. Let $C'(p,q)$ be a dense subset of all the timelike $C^1$ u-curves from $p$ to $q$.

\prop{Lemma 6.7.2} Let $L_c[\lambda]=\int_p^q (-g(\partial/\partial{t},\partial/\partial{t}))^{1/2}dt$. $L_c$ is upper semi-continuous in the $C^0$ topology on $C'(p,q)$.

Define $d_c(p,q)$ for $p,q\in\mathscr{M}_c$ to be zero if $q$ is not in $J^+(p)$ and otherwise to be the least upper bound of the lengths of future-directed piecewise non-spacelike u-curves from $p$ to $q$. For sets $\mathscr{S}$ and $\mathscr{U}$, define $d_c(\mathscr{S},\mathscr{U})$ to be the least upper bound of $d(p,q), p\in\mathscr{S}, q\in\mathscr{U}$:

\prop{Lemma 6.7.3} $d_c(p,q)$ is finite and continuous in $p$ and $q$ when $p$ and $q$ are contained in an c-globally hyperbolic set $\mathscr{N}$.

\prop{Corollary} If $\mathscr{S}$ is a $C^2$ partial c-Cauchy surface, then to each point $q\in{}D^+_c(\mathscr{S})$ there is a future-directed timelike geodesic u-curve orthogonal to $\mathscr{S}$ of length $d_c(\mathscr{S},q)$ which does not contain any point conjugate to $\mathscr{S}$ between $\mathscr{S}$ and $q$.

Any point $p$ in $\mathscr{M}_c$ is uniquely determined by its c-chronological past $I^-_c(p)$ or its c-future $I^+_c(p)$, i.e. $$I^-_c(p)=I^-_c(q)\Longleftrightarrow{}I^+_c(p)=I^+_c(q)\Longleftrightarrow{}p=q.$$
The c-chronological past $\mathscr{W}\equiv{}I^-(p)$ of any point $p\in\mathscr{M}_c$ has the properties: (1) $\mathscr{W}$ is open; (2) $\mathscr{W}$ is a c-past set, i.e. $I^-_c(\mathscr{W})\subset\mathscr{W}$; and (3) $\mathscr{W}$ cannot be expressed as the union of two proper subsets which have the properties (1) and (2).

\prop{Proposition 6.8.1} A set $\mathscr{W}$ is an c-TIP if and only if there is a future-inextendible timelike u-curve $\gamma$ such that $I^-(\gamma)=\mathscr{W}$.

Define $\hat{\mathscr{M}}_c$ as the set of all c-IPs of $(\mathscr{M}_c,\textbf{g})$; $\hat{\mathscr{M}}_c$ is $\mathscr{M}_c$ plus a future causal boundary. Similarly, define $\breve{\mathscr{M}}_c$ as the set of all c-IFs of $(\mathscr{M}_c,\textbf{g})$; $\breve{\mathscr{M}}_c$ is $\mathscr{M}_c$ plus a past causal boundary. The goal is to form a space $\mathscr{M}^*_c$ which has the form $\mathscr{M}_c\cup\Delta$, where $\Delta$ will be called the causal boundary of $(\mathscr{M}_c,\textbf{g})$. Let $\mathscr{M}^\sharp_c=\hat{\mathscr{M}}_c\cup\breve{\mathscr{M}}_c$. To get $\mathscr{M}^*_c$, identify points of $\mathscr{M}^\sharp_c$ to make it Hausdorff.
\bibliography{equivalence}
\end{document}